\documentclass{ws-procs961x669}            % default, citations in superscript
\usepackage{hyperref}
\begin{document}
\title{The Physics of Quantum Information}

\author{John Preskill}

\address{Institute for Quantum Information and Matter, California Institute of Technology\\
AWS Center for Quantum Computing\\
Pasadena, California 91125, USA\\
%$^*$E-mail: preskill@caltech.edu\\
}

\begin{abstract}
Rapid ongoing progress in quantum information science makes this an apt time for a Solvay Conference focused on The Physics of Quantum Information. Here I review four intertwined themes encompassed by this topic: Quantum computer science, quantum hardware, quantum matter, and quantum gravity. Though the time scale for broad practical impact of quantum computation is still uncertain, in the near future we can expect noteworthy progress toward scalable fault-tolerant quantum computing, and discoveries enabled by programmable quantum simulators. In the longer term, controlling highly complex quantum matter will open the door to profound scientific advances and powerful new technologies.

\begin{center}
\textit{
Overview talk at the 28th Solvay Conference on Physics\\
``The Physics of Quantum Information''\\
Brussels, 19-21 May 2022
}
\end{center}
\end{abstract}

%\keywords{needed?}

\bodymatter

%\section{Introduction}

\section{Introduction}
This Solvay Conference on Physics provides a welcome opportunity to assess recent scientific progress and to reflect on the challenges and opportunities before us.  Solvay Conferences had a stirring influence on advances in quantum physics during the 20th century, going back to the very first in 1911. Those advances transformed our understanding of nature, and also led the way to remarkable technologies such as lasers, atomic clocks, magnetic resonance imaging, and billions of transistors on a single microchip. 

Such technologies, though undeniably impressive and impactful, barely scratch the surface of how quantum theory reshapes our view of what’s possible in the universe. Now, for the first time in human history, we are developing and perfecting the tools to create and precisely control very complex states of many interacting particles, states so complex that we cannot efficiently simulate them with our most powerful existing computers or anticipate their behavior using currently known theoretical ideas.  As our ability to control the quantum world matures, profound scientific discoveries and powerful technologies will surely ensue. 

The rapidly unfolding developments in quantum information science make now a particularly apt time for a Solvay Conference on The Physics of Quantum Information. This topic encompasses four intertwined themes to be targeted in subsequent sessions: Quantum computer science,\cite{aaronson-solvay,gottesman-solvay} quantum hardware,\cite{blatt-solvay,schoelkopf-solvay} quantum matter,\cite{verstraete-solvay,lukin-solvay} and quantum gravity.\cite{engelhardt-solvay,stanford-solvay} For each theme I will provide some historical background, then comment on the current status and future prospects.

\section{Background}

\subsection*{Modeling computation}

The fundamental theory of computation builds on foundations erected by Turing in the 1930s.\cite{turing1936computable} Turing defined a computation in terms of an idealized physical process involving manipulation of symbols on a movable tape, and his model became widely accepted as a correct characterization of functions that are in principle computable in the physical world, an assertion known as the Church-Turing thesis. 

A more refined notion, efficient computation, drew attention in the 1970s, igniting the theory of computational complexity.\cite{cook1971complexity,levin1973universal,karp1972reducibility} It became accepted that a problem can be solved efficiently if the number of steps on a Turing machine scales like a polynomial in the size of the input to the problem, the so-called extended Church-Turing thesis. By broad consensus, these are the problems that are feasible to solve in practice. They are said to belong to a complexity class called P, for polynomial time. 

%An example of a problem in P is the palindrome problem, determining whether a string of characters is the same whether read forward or backward. This is easy to solve --- one merely reverses the order of the characters and checks whether the reversed string matches the input string. You can easily check in your head that “Madam in Eden I’m Adam” is a palindrome, if we disregard the apostrophe and the spaces between words. This problem can be solved very quickly by a computer even if the string is billions of characters long. 

Problems in the complexity class NP are those such that a solution, once found, can be efficiently verified by a Turing machine, and it’s generally believed that NP contains hard problems that are outside P. It was noticed that a large family of problems, like combinatorial optimization problems, are in a class called NP-complete – these may be regarded as the hardest problems in the class NP.\cite{cook1971complexity,levin1973universal,karp1972reducibility}

%An example of an NP-complete problem is solving a Sudoku puzzle. In Sudoku one is presented with a partially filled grid, and is challenged to complete the grid by inserting numbers that satisfy specified rules. Humans routinely solve Sudoku on a 9-by-9 grid. Even on a much larger grid, one can very easily check that a candidate solution really satisfies all the rules, so the determining whether a solution exists is a problem in NP. On the other hand, finding a solution rapidly becomes very difficult as the size of the grid increases; it would take a powerful supercomputer far longer than the age of the universe to solve Sudoku if the grid size were, say, 1000-by-1000. We say that Sudoku is NP-hard. 

We believe that NP also contains problems outside P which are not NP-complete. Finding the prime factors of a large composite integer is a famous problem thought to be of this type.  

As a practical application of complexity theory, public key cryptosystems were proposed in the 1970s, based on problems like factoring that are outside P but not NP-complete.\cite{diffie2019new,rivest1978method} These schemes are heavily used today to protect the privacy of electronic communication, and are based on the presumption that a computation that could break the protocol is too hard to carry out in practice.

\subsection*{Quantifying information}

Information theory builds on foundations erected by Shannon in the 1940s.\cite{shannon1948mathematical} Shannon quantified the information conveyed by a message according to how much the message could be compressed to fewer bits without any loss of content. He also quantified how much information can be transmitted from sender to receiver through a noisy communication channel such that the information can be decoded with negligible probability of error by the receiver.

This theory led to the notion of an error-correcting code which can protect redundantly encoded information against the damaging effects of noise;\cite{hamming1950error} this in turn led to results establishing that computations can be performed reliably even when the computing hardware is imperfect.\cite{von1956probabilistic} Error-correcting codes are also vitally important in modern communication systems, like mobile phone cellular networks.

\subsection*{Quantum information}

The origins of quantum information theory can be traced back to observations by Einstein and collaborators in the 1930s,\cite{einstein1935can} who noticed that correlations among parts of a quantum system can have counterintuitive properties, a phenomenon called “quantum entanglement” by Schr\"{o}dinger.\cite{schrodinger1935discussion} John Bell formalized this notion in the 1960s\cite{bell1964einstein} by establishing that players who share quantum entanglement can win a cooperative game with a higher success probability if they share entangled qubits as opposed to classically correlated bits.\cite{clauser1969proposed,aspect1981experimental} In this sense quantum entanglement is a valuable resource that can be consumed to perform useful tasks. 

In the 70s and 80s it was recognized that quantum communication, for example by sending photons through optical fiber or free space, can be advantageous in cryptography, as security can be based on principles of quantum physics rather than limitations on the computational resources available to potential adversaries.\cite{wiesner1983conjugate,bennett2020quantum,ekert1992quantum} The crucial principle is that unknown quantum states, in contrast to classical bits, cannot be copied accurately,\cite{wootters1982single,dieks1982communication} and in fact acquiring information about the content of quantum signals produces an unavoidable disturbance which is in principle detectable. 

Also in the 70s, the general theory of measuring and processing quantum states was developed, including fundamental limits on how much classical information can be acquired when a quantum system is measured.\cite{helstrom1969quantum,holevo1973bounds} 

\subsection*{Quantum computation}

That properties of a complex highly correlated quantum system of many particles are hard to compute is an old observation already known to the pioneers of quantum mechanics. In the early 1980s, Feynman\cite{feynman21simulating} and Manin\cite{manin1980computable} articulated the idea that properties that are hard to calculate with a conventional computer might be easy if we compute with a quantum device instead. This gave rise to a revision of the extended Church-Turing thesis, which in its revised form can be informally stated as, “A quantum computer can efficiently simulate any process that occurs in nature.''\cite{deutsch1985quantum} It is now widely believed, though not proven from first principles, that quantum computers have an exponential advantage over conventional computers for some problems, potentially including  problems of interest in chemistry and materials science. That is, computations that can be performed in a time that scales polynomially with system size using a quantum computer require a time that scales exponentially using a conventional computer. 

It was also found theoretically that quantum algorithms have a superpolynomial advantage over the best known classical algorithms for problems of interest in modern cryptography, such as finding the prime factors of a large composite integer.\cite{simon1997power,shor1999polynomial,kitaev1995quantum} In addition, it is known that quantum computers can speed up exhaustive search for a solution to a combinatorial optimization problem, but in that case the speedup is quadratic, meaning that the quantum time to solution scales like the square root of the classical time.\cite{grover1997quantum,bennett1997strengths} 

%Something about why it is hard to simulate highly entangled quantum systems with a classical computer. The quantum systems speak a different language. 

\subsection*{What is a quantum computer?}

A mathematical model of an ideal quantum computer was formulated, the quantum circuit model, which has these five essential ingredients.\cite{divincenzo2000physical} (1) A physical system harboring many qubits, such that the qubit number can be scaled upward as needed to solve problems of increasing size. (2) The ability to prepare simple standard initial states of the qubits, in effect to cool the system to a state with low entropy. (3) A universal set of entangling quantum operations, called quantum gates, each acting on two or more qubits, universal meaning that by composing many such gates in succession we can approximate arbitrary unitary transformations acting on many qubits. (4) A classical computer that efficiently translates a problem into a suitable circuit of quantum gates. And (5) the ability to measure qubits in a standard basis to read out classical bits providing the result of the computation. The efficiently solvable problems are those that can be solved with high success probability using a number of quantum gates that scales polynomially with the problem’s input size. 

Other physically reasonable models of quantum computation were also studied, such as the topological\cite{freedman2002modular,freedman2002simulation} and adiabatic\cite{farhi2000quantum,aharonov2008adiabatic} models, and shown to be equivalent to the quantum circuit model, thus lending further support to the quantum version of the extended Church-Turing thesis. 

One should appreciate that all the features of the quantum circuit model can be simulated by an conventional classical computer, if equipped by a random number generator to capture the nondeterministic nature of the final quantum measurement. All the classical computer needs to do is keep track of a vector in a Hilbert space as we act on the vector with a sequence of matrices. For the final readout, it projects the vector onto a standard set of axes, and assign probabilities to the different measurement outcomes accordingly. Since a (randomized) classical computer can do whatever a quantum computer does, there is no difference in \emph{computability} --- whatever is computable by a quantum computer is computable by a classical computer as well. 
 
 The important distinction between the quantum and classical models is all about \emph{efficiency}. In general, for the classical computer to simulate the quantum computer, it has to deal with vectors in a space whose dimension is exponential in the number of qubits. For the hardest problem instances, all known classical methods for doing this simulation require resources that scale exponentially with the number of qubits.

%What makes quantum computers more powerful than classical ones? 

%Something about how the notion of computability is not modified. And where does the power come from?

\subsection*{Quantum hardware}

After the discovery of Shor’s factoring algorithm in 1994, interest in quantum computing exploded, igniting pursuit of possible approaches to constructing hardware that could meet the five criteria enumerated above, at least to a reasonable approximation. And it was noticed that some technologies that were already being developed for other reasons could be adapted for the purpose of coherent quantum information processing. 

For example, motivated by the quest for more precise clocks, tools had been developed for cooling and manipulating individual electrically charged atoms using laser fields, which led to ion-trap quantum processors.\cite{cirac1995quantum,monroe1995demonstration} 
%[Monroe, Blatt] 
Josephson junctions, nonlinear elements in superconducting circuits which were being used in high precision magnetometers, led to superconducting quantum processors.\cite{PhysRevB.35.4682,koch2007charge,martinis2002rabi} Experience with nanoscale electrical circuits resulted in the ability to isolate and manipulate spins of single electrons.\cite{PhysRevA.57.120,petta2005coherent} High efficiency sources and detectors for single photons opened the possibility of processors based on photonics.\cite{knill2001scheme} Methods for trapping and cooling neutral atoms led to tunable simulations of strongly interacting quantum matter.\cite{jaksch1998cold,greiner2002quantum} Later, optical tweezers provided an opportunity to build programmable simulators based on arrays of highly excited neutral atoms.\cite{lukin2001dipole,gaetan2009observation,saffman2010quantum,endres2016atom} These and other approaches to quantum hardware are still being developed and are steadily advancing.

%\subsection*{Ions}

Currently the two most advanced quantum computing technologies are ion traps\cite{bruzewicz2019trapped,blatt-solvay} and superconducting circuits.\cite{kjaergaard2020superconducting,schoelkopf-solvay} In an ion trap, each qubit is a single electrically charged atom, which can be in either its ground state or a long-lived excited state. Tens of qubits can be stored in a linear array, and the state preparation, readout, and single-qubit quantum gates can all be achieved by addressing an ion with a stable laser. To perform entangling two-qubit gates, one manipulates the normal modes of vibration of ions in the trap using laser fields --- a two-qubit gate can be executed on any pair of ions in tens of microseconds. 

To scale up to larger systems, one envisions modular trapping regions, connected together by optical interconnects or by shuttling ions from one trapping region to another. 

%Mention current gate fidelities?

%\subsection*{Superconductors}

In a superconducting quantum computer, of order 100 qubits can be arranged in a two-dimensional array, with nearest-neighbor couplings among the qubits. These qubits, called transmons, are in effect artificial atoms that must be carefully fabricated and frequently calibrated. One reads out a transmon by coupling it to a microwave resonator, and single-qubit quantum gates are executed by addressing the qubit with microwave pulses. Two-qubit gates can be performed by various means, for example by tuning the frequencies of a pair of qubits into and out of resonance, or by driving one qubit at the frequency of another. A two-qubit gate takes tens of nanoseconds. 

For scaling up to larger systems, one must address the challenge of dealing with many microwave control lines, and to improve gate fidelities, better materials, fabrication quality, and possibly alternative qubit designs would all be helpful.

%Fidelities? Tunable couplers?

By now, quantum processors have advanced to the stage where they can perform tasks that are challenging to simulate using classical computers. In particular, one can sample from the output probability distribution of randomly chosen circuits with 60 qubits and over 20 cycles of entangling two-qubit gates.\cite{arute2019quantum,wu2021strong,zhu2022quantum} Though this specific task is not of intrinsic practical interest, such experiments have been useful, by providing new benchmarks for circuit fidelity, solidifying our understanding of the global features of circuit noise, and provoking improvements in classical simulation methods.\cite{pan2021solving}

\subsection*{Quantum error correction}

When Shor’s algorithm was discovered, and interest in quantum computing surged, there was widespread and understandable skepticism regarding whether large-scale quantum computing would ever be practical.\cite{unruh1995maintaining,landauer1995quantum,haroche1996quantum} Quantum systems have the inconvenient property that observing a quantum state inevitably disturbs the state in an uncontrolled way, so interactions with the environment cause quantum information to decay rapidly, a phenomenon called decoherence. To execute a quantum computation reliably we must keep the information we process nearly perfectly isolated from the outside world to prevent decoherence, which is quite difficult because our hardware can never be perfect. 

It was quickly discovered that, at least in principle, hardware imperfections can be overcome with suitable software based on what we call quantum error-correcting codes.\cite{shor1995scheme,steane1996error,knill1997theory,gottesman1997stabilizer} The crucial idea is that we can protect quantum information by storing it nonlocally, encoding it in a very highly entangled form such that when the environment interacts with the parts of the system locally, it acquires negligible information about the encoded quantum state and so need not damage the state. Furthermore, we learned how to process efficiently quantum information that is encoded in this highly entangled way.\cite{shor1996fault} It follows that, if errors in a quantum computer are sufficiently rare and not too strongly correlated, we can simulate an ideal quantum computation efficiently using a noisy quantum computer.\cite{aharonov2008fault,knill1998resilient,kitaev1997quantum,preskill1998reliable,preskill1998fault,aliferis2005quantum,reichardt2006fault} 

The most promising protocol for error-corrected quantum computing in the relatively near term is based on Kitaev’s surface code,\cite{kitaev2003fault} which has two advantages: it can tolerate a relatively high physical error rate,\cite{dennis2002topological,raussendorf2007fault,fowler2012surface} and it requires only geometrically local processing in a two-dimensional layout. Even so, the overhead cost of error correction, in both the number of physical qubits needed and the number of physical gates, is quite daunting. One can plausibly anticipate running algorithms for a few hundred protected logical qubits that would surpass the best conventional computers for some problems of practical interest, but to achieve sufficiently reliability the number of physical qubits might be in the millions. That's a big leap from the devices we expect to have in the next few years, with hundreds of physical qubits.

% Say something about the concept of quantum error correction --- it makes use of entanglement to hide quantum information. That is already here, but could be amplified. Comment on approximate QEC? Comment on noise mitigation?

\subsection*{Quantum matter}

Deep connections between quantum information and quantum matter emerged with the discovery of topological order\cite{wen1990topological} (initially in fractional quantum Hall systems\cite{tsui1982two,laughlin1983anomalous}), which we now recognize as a manifestation of long-range entanglement in a quantum phase of matter.\cite{chen2010local} By long-range entanglement, we mean that the time needed to prepare the quantum phase, using spatially local operations in a quantum computer and starting with an unentangled state, scales with the total size of the system. Furthermore, topologically ordered phases of matter may be fruitfully viewed as quantum error-correcting codes which conceal nonlocally encoded quantum information.\cite{kitaev2003fault} Symmetry-protected topological phases were also discovered,\cite{haldane1983nonlinear,kane2005quantum,chen2012symmetry} for which the time to prepare the quantum state scales with system size if all local operations in the quantum circuit are required to satisfy specified symmetries. 

It was discovered that ground states of quantum systems often obey an entanglement “area law,” meaning that the amount of entanglement between the particles inside and outside of a specified ball-shaped region scales not like the total number of particles in the region, but rather like the number of particles near the boundary of the region.\cite{bombelli1986quantum,srednicki1993entropy,hastings2007area} This led to new methods for simulating quantum many-body systems on classical computers based on tensor networks, which exploit this entanglement structure to improve substantially on previous methods.\cite{white1992density,fannes1992finitely,vidal2004efficient,verstraete2004renormalization} And it was noticed that entanglement, when quantified by the entropy of the marginal quantum state for the particles in a region, has universal properties that can be used to identify distinct quantum phases of matter.\cite{kitaev2006topological,levin2006detecting,li2008entanglement} 

The computational hardness of preparing quantum ground states was studied, and it was argued convincingly that in some cases this is a hard problem for quantum computers;\cite{kitaev2002classical} this hardness can persist even for translationally invariant one-dimensional systems,\cite{gottesman2009quantum} though admittedly such computationally intractable quantum many-body systems may have exotic interactions that are not necessarily of practical physical interest. At any rate, according to the quantum extended Church-Turing thesis, these ground states that are intractable for quantum computers could not arise in nature by any feasible physical process. 

Quantum information has also provided a fresh perspective on the behavior of strongly chaotic quantum systems, which we now view through the lens of entanglement dynamics.\cite{calabrese2005evolution} Information that is imprinted locally on a quantum system quickly spreads, becoming encoded in the form of quantum entanglement shared by many particles, and hence invisible to local observers who have access to just a few particles at a time. This entanglement spreading can be efficiently simulated by quantum computers,\cite{lloyd1996universal} but is beyond the reach of known classical computational methods, which cannot succinctly encode or efficiently simulate highly entangled many-particle quantum states. 

%What about Lieb Robinson bounds? Hastings? Chaos and ETH? And .. topological phases obey the QEC conditions. They are codes. Mention here that quantum simulators like ultracold atoms in optical lattices can imitate quantum matter. 

\subsection*{Quantum gravity}

The connection between quantum gravity and quantum information can be traced back to Hawking’s 1974 discovery that due to quantum effects black holes emit thermal radiation, arising because of quantum entanglement between the inside and outside of the black hole’s event horizon.\cite{Hawking1975} This led to a quantitative relationship between the area of the event horizon and the black hole’s entropy,\cite{bekenstein1974generalized} a measure of how much quantum information the black hole can store. These results anticipated the area law for entanglement entropy in condensed matter physics that would be discovered years later. 
Furthermore, the entropy of a black hole is astonishingly large --- for example, the entropy of a solar mass black hole, which is just a few kilometers across, is 20 orders of magnitude larger than the entropy of the sun. Indeed black holes, though remarkably simple objects as described by classical gravitation theory,\cite{israel1967event,carter1971axisymmetric} are quantum-mechanically the most complex objects nature allows, as quantified by the black hole's information storage capacity. 

Holographic duality, discovered in the 1990s, established that, at least in negatively curved anti-de Sitter space, quantum gravity in bulk quantum spacetime is equivalent to a nongravitational quantum field theory in one lower dimension that resides on the boundary of the spacetime.\cite{maldacena1999large} And it turned out that the bulk geometry is encoded on the boundary by the structure of the quantum entanglement in the boundary theory.\cite{ryu2006holographic} Furthermore, the holographic dictionary which maps local bulk observables to the corresponding highly nonlocal observables on the boundary was recognized as the encoding map of a kind of quantum error-correcting code.\cite{almheiri2015bulk,pastawski2015holographic} So we can regard the geometry of spacetime itself as an emergent feature arising from underlying quantum entanglement, which is intrinsically robust with respect to some deformations of the boundary theory. 

The entanglement dynamics of black holes was studied, and it was conjectured that black holes are the most efficient scramblers of quantum information that nature will allow.\cite{hayden2007black,sekino2008fast} Here, too, studies of information scrambling that originated in studies of black hole physics stirred growing interest in how information becomes scrambled in other quantum many-body systems that are more accessible in the laboratory.\cite{shenker2014black,maldacena2016bound,cotler2017black}

The entropy of the Hawking radiation emitted by an evaporating black hole, which tracks the evolving quantum entanglement of the radiation with the hole, was studied quantitatively,\cite{almheiri2020replica,penington2022replica} and calculations confirmed that the entropy evolves as expected if the evaporation process is correctly described by unitary quantum theory.\cite{page1993information} Rather unexpectedly, this unitary behavior can be captured by semiclassical computations without reference to the microscopic details of quantum gravity. Such results indicate that black hole physics is profoundly nonlocal; one can in principle access the black hole interior by manipulating radiation that is far away, but only by performing quantum operations which are of such high computational complexity as to be infeasible in practice.\cite{harlow2013quantum} 

%Or just Susskind, to cover complexity as well. Mention complexity of the dictionary, Harlow-Hayden, Hayden-Preskill? Highlight the progress over the past three years. A new entropy formula, for fine-grained entropy. 

%The idea that a falling object is dual to the spreading of a perturbation in a chaotic quantum system. 

\subsection*{Connections}

The discussion so far has already illustrated the many cross-connections among the scientific themes that are represented at this conference. For example, information scrambling is now studied in quantum computing circuits, in chaotic many-particle systems, and in black holes. Quantum error correction, introduced for the purpose of extending quantum computing to large systems, is also relevant to topological phases of matter, and to the holographic correspondence in quantum gravity. Computational complexity, the study of the hardness of computational problems, turns out to be relevant to preparation of topological quantum phases of matter, and also to the geometry of the black hole interior. These are a few examples among many such connections.

\section{Status and prospects}

\subsection*{Where are we now?}

Coming back to quantum computing technology, what is the status today? There are two central questions about quantum computing, which could already have been articulated 40 years ago. How will we scale up to quantum computing systems that can solve hard problems? And how can we best make use of that computational power in science and in industry? In my view, both questions are wide open. How we direct our effort should be guided by that realization.

%\subsection*{What to do with near-term quantum computers?}

One may ask, what should we do with the noisy intermediate-scale quantum computers that we have now?\cite{preskill2018quantum} Two obvious answers are: We should use near-term quantum computers to learn how to build more powerful quantum computers that can have a practical impact. And we should seek a clearer understanding of how those practical quantum computers can eventually be used. 

%\subsection*{Prospects for the next 5 years}

Even if broadly useful quantum computers are still a ways off, much can be accomplished over the next five years or so. In that time frame we can expect to see encouraging progress toward scalable fault-tolerant quantum computing. And we can anticipate scientific discoveries enabled by programmable quantum simulators and circuit-based quantum computers. 

\subsection*{Progress toward quantum error correction}
What would constitute notable progress toward fault-tolerant quantum computing? %Here are some of the capabilities we need. 
We need to be able to do repeated rounds of accurate error syndrome measurement for quantum error correction. And we would like to see concrete evidence that quantum memory times continue to improve sharply as we include more and more physical qubits to encode each protected logical qubit.

The ion trappers could justifiably protest that they don’t care very much about quantum memory times, because their atomic qubits already have extraordinarily long lifetimes. That’s true. But for all currently foreseen platforms it is crucial to achieve much higher fidelity for entangling two-qubit logical gates --- only then will we be able to run powerful quantum algorithms. What we may hope to see in the near term are logical two-qubit gates, protected by quantum error correction, with much higher fidelity than our best physical two-qubit gates, as well as solid evidence that logical gate fidelities continue to improve sharply as the code block increases in size. That has not yet been accomplished. 

%\subsection*{Quantum error correction circa 2022}

What is the status now? There has been exciting recent progress toward quantum error correction; I’ll highlight two contributions, one from Google and one from Honeywell (now called Quantinuum).\footnote{This discussion reflects progress that had already been reported at the time of the Solvay Conference in May 2022.} 

Google investigated the quantum repetition code, using up to 21 qubits in the Sycamore processor, 11 in the code block and 10 ancilla qubits for error syndrome readout.\cite{ai2021exponential} Importantly, this is not a full-fledged quantum error-correcting code --- it protects against dephasing errors but not against bit flips. Nonetheless, it was an impressive demonstration. They did up to 50 consecutive rounds of syndrome measurement, each taking about 1 microsecond, with most of that time devoted to resetting the ancilla qubits to prepare for the next round of syndrome measurement. They observed that the rate of logical errors due to dephasing decreased by about a factor of 10 each time the code distance increased by 4, for example as the code length increased from 3 to 7 and from 7 to 11. That was in accord with expectations given the noise in their device. 

Quantinuum demonstrated error correction for a 7-qubit code that can correct an arbitrary error acting on any one qubit out of the 7.\cite{ryan2021realization} They did up to 6 consecutive rounds of error correction, each taking 200 milliseconds. Note the much different cycle times for the superconducting and ion trap devices. As quantum computing advances, and the time on the wall clock for running an algorithm becomes an increasingly important consideration, that difference may loom large. Quantinuum uses an architecture in which ions are transported to processing zones where rather high fidelity operations can be performed in parallel, and after movement they use another ion species to sympathetically cool the motional state. That cooling enables them to do the repeated rounds of syndrome measurement, but also accounts for much of the time budget of their circuit. 

Unfortunately, gate error rates in the Google and Honeywell machines, and other current devices,\cite{egan2020fault,krinner2022realizing} are still too high for quantum error correction to improve two-qubit logical gate fidelities. 

%IBM, Innsbruck, etc. Google again.\cite{acharya2022suppressing} Quantinuum again.\cite{ryan2022implementing}

\subsection*{Fault-tolerance with the surface code}

%Where do we go from here? 
The best currently known prospect for scaling up quantum computing in the relatively near term is based on the surface code, introduced by Alexei Kitaev over 25 years ago.\cite{kitaev2003fault} As already noted, the two great virtues of the surface code are that error syndromes can be extracted using only geometrically local processing in a two-dimensional layout, and that each syndrome bit can be read out using a simple quantum circuit involving only four data qubits. As a result, the surface code can tolerate higher error rates than other feasible quantum codes.\cite{dennis2002topological,raussendorf2007fault,fowler2012surface} 

Despite being more effective than other codes, error correction with the surface code still carries a rather hefty overhead cost in the number of qubits and gates needed. Let’s suppose we can do physical controlled-NOT gates with an error rate of .1\%. That’s better than we have now in multiqubit devices, but might plausibly be reached in the near future. Perhaps we will start to see quantum advantage %, for example in simulations of materials, 
using circuits with hundreds of protected qubits and millions of high-fidelity quantum (Toffoli) gates. To execute those circuits, we’re likely to need at least tens of thousands of physical qubits. For breaking public key cryptosystems using Shor’s algorithm, it is estimated that 20 million physical qubits are needed.\cite{gidney2021factor} If we can somehow do controlled-NOT gates with four 9s of fidelity, an improvement by another order of magnitude, that will reduce the overhead cost, but we’re still likely to want at least hundreds of physical qubits per logical qubit to see significant quantum advantage running algorithms we currently know about. These numbers are surely daunting from the perspective of currently available technology.

In an exciting recent development, quantum codes have been discovered that are far more efficient than the surface code.\cite{gottesman-solvay,panteleev2022asymptotically,leverrier2022quantum} Someday we might use these codes to reduce significantly the overhead cost of fault-tolerant quantum computing. 
As best we currently understand, though, to perform well these codes require much lower physical error rates than the surface code, and so are not likely to be useful until much better quantum hardware is available. 

\subsection*{Much better gate error rates?}

It would pay off handsomely to have much improved physical gate error rates in quantum hardware, but that’s very hard to achieve. A particularly visionary proposal is topological quantum computing, where qubits are encoded in an exotic material that provides physical protection against noise.\cite{kitaev2001unpaired} High fidelity topologically protected quantum gates, if realized, would be a genuine milestone for quantum many-body physics, aside from any implications for future information technologies. Though the theoretical idea is compelling, up to now experimental progress has been slow.\cite{aghaee2022inas}

There are other potential ways to incorporate better protection against noise into the hardware itself. Some promising ideas exploit precise manipulation of bosonic modes, such as microwave resonators in superconducting circuits, harmonic motion of trapped ions, or optical modes in photonic devices. For example, GKP-encoded states of bosonic modes have a periodic grid structure in phase space, allowing slight shifts in phase space to be corrected.\cite{gottesman2001encoding,campagne2020quantum,fluhmann2019encoding} Bosonic cat codes use superpositions of coherent states to provide strong protection against bit flips, resulting in highly biased physical noise that can be corrected by quantum codes at a reduced overhead cost.\cite{puri2019stabilized,lescanne2020exponential,grimm2020stabilization} Fluxonium qubits\cite{nguyen2019high,bao2022fluxonium} and zero-pi qubits\cite{brooks2013protected,gyenis2021experimental} use strong nonlinearity resulting from large inductance in a superconducting circuit to suppress noise. In the arena of superconducting qubits, all of these schemes are more complex than the relatively simple transmon; they are still at a comparatively early stage, and we can't say yet how they'll pan out. But it is important to continue pursuing these and other challenging approaches offering the potential for a leap forward in performance, because significantly lower physical gate error rates will bring us closer to useful applications of quantum computation.  

%Part of making those improvements involves materials issues. That takes a lot of time and resources and the time scale for progress is difficult to predict. At any rate, we need more than just better materials, we need innovative ideas about how to make use of those materials for better qubit designs. 

\subsection*{Creating quantum states of matter}

The quantum technology we already have is exciting, as it provides new tools for exploring the physics of many entangled particles. On this front, too, we’ve seen significant recent progress, including unprecedented studies of new quantum phases of matter. I’ll highlight two examples. 

The Harvard/MIT group, using a Rydberg atom platform with 219 qubits, recently created and detected a novel highly entangled phase of quantum matter, a quantum spin liquid.\cite{semeghini2021probing} Theorists have speculated about quantum spin liquids for nearly 50 years,\cite{anderson1973resonating} but convincing experimental evidence for this type of quantum state had never been seen before, for two main reasons. First, one needs a material with suitable properties for qubits to seek a ground state with long-range quantum entanglement. In nature, such materials seem to be rare. Second, the features of a long-range-entangled state are very elusive to observe because one needs to make collective observations on many qubits at once. The Rydberg platform is highly programmable and sufficiently versatile to simulate the right kind of material. And one can measure nonlocal observables with sufficient fidelity to identify signatures of long-range entanglement.

Guided by university condensed matter physicists from Stanford, Princeton, the Max Planck Institute and elsewhere, 20 superconducting qubits in the Google Sycamore processor were used to create and observe a discrete time crystal.\cite{mi2022time} This is a novel phase of matter in a periodically driven system, which oscillates indefinitely at a frequency different from that of the periodic drive. The idea of a time crystal had been suggested for the first time 10 years ago,\cite{wilczek2012quantum} and there had been previous experiments which were partially successful in validating the phenomenon,\cite{choi2017observation,zhang2017observation,kyprianidis2021observation} but the high fidelity gates and accurate single-qubit readout and control in Sycamore made a more convincing demonstration possible. 

Take note of two things. First, five years ago Rydberg atoms were not so much on the radar screen of quantum platforms, yet now they are advancing rapidly. That reminds us that we are still in the early days of quantum technology, and big surprises may continue to arise. Second, the Google experiment was done on a gate-based quantum computer, while the Harvard/MIT experiment was done in a programmable analog mode. That reminds us that these two approaches to studying quantum matter are complementary, and both are valuable to pursue. 

These are encouraging signs that we’re acquiring tools to create and investigate a variety of other new quantum phases of matter in the near future, both in equilibrium like a quantum spin liquid, or driven far from equilibrium like a discrete time crystal. There are good reasons to be impressed by these developments. First because, of the applications of quantum computing that we currently foresee, those to materials and chemistry are the ones that seem to have the greatest potential to benefit humanity broadly speaking, and it’s exciting that we already have tools in the current era that can advance our understanding of quantum matter. And second because studies of topological phases could ignite new approaches to quantum error correction and fault tolerance that will pay off in the longer run. Looking ahead, we can glimpse opportunities to create states of matter beyond what is known to occur in nature, which can have both scientific and technological value. 

%Comment on other experiments.\cite{chiu2019string,abanin2019colloquium} [Greiner, Bloch] We’re not limited to simulating the materials that occur in nature. We can do much more than that. 

\subsection*{Opportunities in quantum simulation}

What will be the long term impact of quantum computing on society? No one knows that. Nor should we be expected to envision clearly how quantum computing might change the world. As I’ve said, of the applications we currently most clearly foresee, what we imagine is most likely to broadly benefit humanity are applications to chemistry and materials, which could improve human health, energy production, agriculture, and the sustainability of our planet. 

Can we be more concrete about that expected impact? That’s quite difficult for several reasons. We seek applications of quantum computing that meet three criteria. The problems of interest should be too hard to solve with conventional computing, efficiently solvable by quantum computers, and the solutions should be of scientific and/or practical value. 

There are methods for simulating complex molecules and highly correlated materials using conventional computers which are actually rather good, and getting better fast, not just because conventional computers are becoming more powerful, but even more importantly because the classical algorithms are getting better.\cite{verstraete-solvay} For computing properties of ground states and other low-energy states, the classical methods are heuristic, without rigorous performance guarantees. But numerical evidence suggests that the resources needed to obtain accurate results using classical methods such as those based on tensor networks and neural networks scale reasonably with system size for typical molecules or materials that are of scientific interest, because these systems are not so profoundly entangled. If that’s true, the advantage enjoyed by quantum computers may scale polynomially rather that exponentially for such problems.\cite{lee2022there} And the competing quantum methods are also heuristic, because to obtain accurate results efficiently we must be able to prepare states in the quantum computer that have a substantial overlap with the targeted quantum states, which is not rigorously guaranteed. The general purpose method for performing the state preparation task is the adiabatic method, which can be very expensive in systems where there are multiple competing phases separated by first-order phase transitions, as is often true in cases of interest. 

Exponential quantum advantage can be expected in quantum simulation of dynamics, if we consider easily prepared initial excited states which become highly entangled as they evolve, for example in highly inelastic collisions of fundamental particles in a quantum field theory.\cite{jordan2012quantum} What scientific opportunities that may entail is an issue worthy of further investigation.

%What references here? Comment further on opportunities in dynamics.\cite{daley2022practical} The idea that simulating quantum chaos will open new opportunities, as did simulations of classical chaos. Analog simulators – how accurate do they need to be. How robust are the properties of interest?

\subsection*{Challenges in quantum gravity}

Circling back to quantum gravity, what are some of the challenges where we can realistically expect to make substantial progress in the reasonably near future? 

In the case of quantum gravity in anti-de Sitter space, we still lack a good handle on why local quantum physics provides an excellent approximation on distance scales that are small compared to the scale of the spatial curvature. In addition, the spacetime we live in is not anti-de Sitter, and we need better tools for describing quantum gravity in spacetimes that are asymptotically flat or positively curved. Anti-de Sitter space has the convenient feature that spacetime has a boundary, and we can define the observables of the theory by making reference to that boundary. But de Sitter space, which is relevant to early-universe inflationary cosmology, does not have that convenient feature, which makes quantum gravity in de Sitter space intrinsically harder to think about. 

Despite remarkable recent progress,\cite{engelhardt-solvay,stanford-solvay} we don’t have an adequate way, in quantum gravity, to describe the experience of an observer who falls into a black hole, and we especially don’t know what happens to observers who encounter the singularity in the black hole interior. 

Holographic duality is very empowering, but we have analytic control over how it works only in a limited number of special cases. Can we understand more systematically under what conditions a nongravitational boundary theory admits a holographic dual which is useful for describing quantum gravitational phenomena?

And can we learn more concretely what resources we’ll need to simulate quantum gravity with quantum computers, and compute observable properties of scientific interest?

\subsection*{Quantum gravity: can experiments help?}

Eventually, we might hope to make progress on some of these questions by using quantum computers and quantum simulators; in particular, by simulating strongly-coupled quantum many-body systems and leveraging holographic duality, we might probe the dual quantum geometry by measuring features of quantum entanglement on the boundary. We might, for example, learn about locality in the bulk spacetime though linear response measurements that yield information about the commutators of boundary observables. Studying the entanglement dynamics of strongly chaotic systems can reveal how quantum information gets scrambled, which might unveil signatures of string theory in the bulk. Or we might in other contexts be able to measure quantum corrections to semiclassical gravity that would be hard to compute analytically or by using classical computers. Simulations of very-high-energy scattering in the bulk could be especially instructive. 

Perhaps guidance from simulations can help us to grasp holographic dual descriptions of spacetimes beyond anti-de Sitter space. And we may find that some otherwise opaque features of strongly-coupled dynamics can be more easily interpreted using the lens of bulk quantum gravity. One already much studied example is a mysterious type of coherent quantum teleportation in the boundary theory which has a quite natural alternative interpretation in terms of quantum information transmitted through a traversable spatial wormhole in the bulk theory.\cite{gao2017traversable,maldacena2017diving,brown2019quantum,nezami2021quantum}

%We can do experiments which probe the high energy behavior of the bulk theory. That could teach us something new about quantum gravity.\cite{shenker2014black} 

\section{Some things I haven’t mentioned}

There are some important things I have not yet had time to mention in this talk, of which I list four here. 

The discovery of Shor’s algorithm will have a disruptive effect on electronic commerce, as the public key cryptosystems we now rely on to protect our privacy will no longer be secure when powerful quantum computers are readily available. The world is responding by developing new classical cryptosystems that are widely believed to be resistant to attack by quantum computers.\cite{bernstein2017post} It will be a necessary task, but a prolonged and expensive one, to deploy these new systems. 

An alternative approach to protecting our privacy is to distribute secure private keys through quantum communication, presumably by sending photons through optical fiber or free space.\cite{bennett2020quantum,ekert1992quantum} Here the security rests on principles of quantum physics, rather than assumptions about the computational power of our adversaries, and in fact there are protocols which are provably secure even if we don’t trust the equipment we use to distribute the key.\cite{vazirani2019fully} It’s not clear to what extent the world will demand quantum cryptography for secure communication; in any event, quantum key distribution on a global scale will require new technologies which are now nascent, like quantum repeaters to extend the range of quantum communication, which in turn are likely to rely on transduction of single-photon signals from optical to microwave frequencies and back.\cite{wehner2018quantum} As is the case for quantum computing, we still lack a clear understanding of what will be the most impactful future applications of quantum networking.

Advancing quantum technology will also enhance the sensitivity and resolution of sensors, which can be expected to have widespread applications including inertial sensors for navigation, gravity gradiometers for surveying, magnetometers for noninvasive nanoscale imaging of living mater, and many others.\cite{degen2017quantum} There are also applications of fundamental interest, including looking for symmetry violations in searches for new physics, dark matter detection, detection of gravitational waves with enhanced sensitivity, and long-baseline optical interferometry enabled by quantum teleportation within a network of telescopes. These improvements will be based on advanced quantum strategies, which exploit squeezing, entanglement, and quantum error correction. 

Another important issue is: how can we be sure that a quantum computation gives the right answer? In some cases, like when factoring a large number, the answer once found can be easily checked with a classical computer, but that’s not the case if, for example, we are simulating the properties of a complex quantum many-body system. Yet clever protocols have been developed for verifying that a quantum computer really performs an assigned task; these leverage the power of quantum-resistant cryptography.\cite{mahadev2018classical} One important challenge is to reduce the cost of these verification protocols so we can make use of them in the relatively near term, if for example we send a job to a quantum server in the cloud, and want to be confident that the answer we receive can be trusted. 

\section{Conclusions}

To conclude, we might have a long road ahead to practical commercial applications of quantum computing, and quantum error correction is most likely the key to getting there eventually.  But the next five years should be exciting, marked by progress toward fault-tolerant quantum computing and unprecedented opportunities to explore exotic properties of quantum matter. 

%Quantum sensing will advance together with quantum networking and quantum computing, and will enable both impactful practical applications and new tools for exploring fundamental physics.  

As this conference has illustrated, The Physics of Quantum Information provides unifying concepts and powerful technologies for controlling and exploring complex many-particle quantum systems of both practical and fundamental interest. 
%We sometimes worry that as science progresses, it continually splinters into narrower and narrower specialties that interact less and less with one another. But there is a countervailing trend: As knowledge advances, scientists working on different topics find that they have more and more to learn from one another.
%
Communication among the practitioners of quantum computer science, quantum hardware, quantum matter, and quantum gravity %benefits us all, 
sparks new ideas and insights, making life sweeter for all of us who investigate the elusive properties of highly entangled quantum systems. 

For the longer term quantum science and technology face enormous challenges, and many advances in both basic research and systems engineering will be needed to fulfill our aspirations. We’ve only just begun. 

\bibliographystyle{ws-procs961x669}
\bibliography{preskill}

\begin{thebibliography}{100}

\bibitem{aaronson-solvay}
S.~Aaronson, How much structure is needed for huge quantum speedups?, these
  proceedings.

\bibitem{gottesman-solvay}
D.~Gottesman, Opportunities and challendges in fault-tolerant quantum
  computation, these proceedings.

\bibitem{blatt-solvay}
R.~Blatt, The trapped-ion platform for quantum information processing, these
  proceedings.

\bibitem{schoelkopf-solvay}
R.~Schoelkopf, Superconducting qubits as a platform for quantum computation,
  these proceedings.

\bibitem{verstraete-solvay}
F.~Verstraete, Quantum information and many-body systems, these proceedings.

\bibitem{lukin-solvay}
M.~Lukin, Programmable quantum machines for probing entanglement in many-body
  systems, these proceedings.

\bibitem{engelhardt-solvay}
N.~Engelhardt, The entropy of {Hawking} radiation, these proceedings.

\bibitem{stanford-solvay}
D.~Stanford, Quantum information and spacetime, these proceedings.

\bibitem{turing1936computable}
A.~M. Turing {\em et~al.}, On computable numbers, with an application to the
  entscheidungsproblem, {\em J. of Math} {\bf 58}, p.~5  (1936).

\bibitem{cook1971complexity}
S.~A. Cook, The complexity of theorem-proving procedures, in {\em Proceedings
  of the third annual ACM Symposium on Theory of Computing\/},  1971 pp.
  151--158.

\bibitem{levin1973universal}
L.~A. Levin, Universal sequential search problems, {\em Problemy peredachi
  informatsii} {\bf 9}, 115  (1973).

\bibitem{karp1972reducibility}
R.~M. Karp, Reducibility among combinatorial problems, in {\em Complexity of
  Computer Computations\/},  (Springer, 1972) pp. 85--103.

\bibitem{diffie2019new}
W.~Diffie and M.~E. Hellman, New directions in cryptography, in {\em Secure
  Communications and Asymmetric Cryptosystems\/},  (Routledge, 2019) pp.
  143--180.

\bibitem{rivest1978method}
R.~L. Rivest, A.~Shamir and L.~Adleman, A method for obtaining digital
  signatures and public-key cryptosystems, {\em Communications of the ACM} {\bf
  21}, 120  (1978).

\bibitem{shannon1948mathematical}
C.~E. Shannon, A mathematical theory of communication, {\em The Bell System
  Technical Journal} {\bf 27}, 379  (1948).

\bibitem{hamming1950error}
R.~W. Hamming, Error detecting and error correcting codes, {\em The Bell System
  Technical Journal} {\bf 29}, 147  (1950).

\bibitem{von1956probabilistic}
J.~Von~Neumann, Probabilistic logics and the synthesis of reliable organisms
  from unreliable components, {\em Automata Studies} {\bf 34}, 43  (1956).

\bibitem{einstein1935can}
A.~Einstein, B.~Podolsky and N.~Rosen, Can quantum-mechanical description of
  physical reality be considered complete?, {\em Physical Review} {\bf 47}, p.
  777  (1935).

\bibitem{schrodinger1935discussion}
E.~Schr{\"o}dinger, Discussion of probability relations between separated
  systems, in {\em Mathematical Proceedings of the Cambridge Philosophical
  Society\/},  (4) 1935 pp. 555--563.

\bibitem{bell1964einstein}
J.~S. Bell, On the {Einstein Podolsky Rosen} paradox, {\em Physics Physique
  Fizika} {\bf 1}, p. 195  (1964).

\bibitem{clauser1969proposed}
J.~F. Clauser, M.~A. Horne, A.~Shimony and R.~A. Holt, Proposed experiment to
  test local hidden-variable theories, {\em Physical Review Letters} {\bf 23},
  p. 880  (1969).

\bibitem{aspect1981experimental}
A.~Aspect, P.~Grangier and G.~Roger, Experimental tests of realistic local
  theories via {Bell's} theorem, {\em Physical Review Letters} {\bf 47}, p. 460
   (1981).

\bibitem{wiesner1983conjugate}
S.~Wiesner, Conjugate coding, {\em ACM Sigact News} {\bf 15}, 78  (1983).

\bibitem{bennett2020quantum}
C.~H. Bennett and G.~Brassard, Quantum cryptography: Public key distribution
  and coin tossing, {\em arXiv preprint arXiv:2003.06557}   (2020).

\bibitem{ekert1992quantum}
A.~K. Ekert, Quantum cryptography and {Bell’s} theorem, in {\em Quantum
  Measurements in Optics\/},  (Springer, 1992) pp. 413--418.

\bibitem{wootters1982single}
W.~K. Wootters and W.~H. Zurek, A single quantum cannot be cloned, {\em Nature}
  {\bf 299}, 802  (1982).

\bibitem{dieks1982communication}
D.~Dieks, Communication by {EPR} devices, {\em Physics Letters A} {\bf 92}, 271
   (1982).

\bibitem{helstrom1969quantum}
C.~W. Helstrom, Quantum detection and estimation theory, {\em Journal of
  Statistical Physics} {\bf 1}, 231  (1969).

\bibitem{holevo1973bounds}
A.~S. Holevo, Bounds for the quantity of information transmitted by a quantum
  communication channel, {\em Problemy Peredachi Informatsii} {\bf 9}, 3
  (1973).

\bibitem{feynman21simulating}
R.~P. Feynman, Simulating physics with computers, 1981, {\em International
  Journal of Theoretical Physics} {\bf 21}.

\bibitem{manin1980computable}
Y.~Manin, Computable and {Uncomputable}, {\em Sovetskoye Radio, Moscow} {\bf
  128}  (1980).

\bibitem{deutsch1985quantum}
D.~Deutsch, Quantum theory, the {Church--Turing} principle and the universal
  quantum computer, {\em Proceedings of the Royal Society of London. A.
  Mathematical and Physical Sciences} {\bf 400}, 97  (1985).

\bibitem{simon1997power}
D.~R. Simon, On the power of quantum computation, {\em SIAM Journal on
  Computing} {\bf 26}, 1474  (1997).

\bibitem{shor1999polynomial}
P.~W. Shor, Polynomial-time algorithms for prime factorization and discrete
  logarithms on a quantum computer, {\em SIAM Review} {\bf 41}, 303  (1999).

\bibitem{kitaev1995quantum}
A.~Y. Kitaev, Quantum measurements and the abelian stabilizer problem, {\em
  arXiv preprint quant-ph/9511026}   (1995).

\bibitem{grover1997quantum}
L.~K. Grover, Quantum mechanics helps in searching for a needle in a haystack,
  {\em Physical Review Letters} {\bf 79}, p. 325  (1997).

\bibitem{bennett1997strengths}
C.~H. Bennett, E.~Bernstein, G.~Brassard and U.~Vazirani, Strengths and
  weaknesses of quantum computing, {\em SIAM Journal on Computing} {\bf 26},
  1510  (1997).

\bibitem{divincenzo2000physical}
D.~P. DiVincenzo, The physical implementation of quantum computation, {\em
  Fortschritte der Physik: Progress of Physics} {\bf 48}, 771  (2000).

\bibitem{freedman2002modular}
M.~H. Freedman, M.~Larsen and Z.~Wang, A modular functor which is universal for
  quantum computation, {\em Communications in Mathematical Physics} {\bf 227},
  605  (2002).

\bibitem{freedman2002simulation}
M.~H. Freedman, A.~Kitaev and Z.~Wang, Simulation of topological field theories
  by quantum computers, {\em Communications in Mathematical Physics} {\bf 227},
  587  (2002).

\bibitem{farhi2000quantum}
E.~Farhi, J.~Goldstone, S.~Gutmann and M.~Sipser, Quantum computation by
  adiabatic evolution, {\em arXiv preprint quant-ph/0001106}   (2000).

\bibitem{aharonov2008adiabatic}
D.~Aharonov, W.~Van~Dam, J.~Kempe, Z.~Landau, S.~Lloyd and O.~Regev, Adiabatic
  quantum computation is equivalent to standard quantum computation, {\em SIAM
  Review} {\bf 50}, 755  (2008).

\bibitem{cirac1995quantum}
J.~I. Cirac and P.~Zoller, Quantum computations with cold trapped ions, {\em
  Physical Review Letters} {\bf 74}, p. 4091  (1995).

\bibitem{monroe1995demonstration}
C.~Monroe, D.~M. Meekhof, B.~E. King, W.~M. Itano and D.~J. Wineland,
  Demonstration of a fundamental quantum logic gate, {\em Physical Review
  Letters} {\bf 75}, p. 4714  (1995).

\bibitem{PhysRevB.35.4682}
J.~M. Martinis, M.~H. Devoret and J.~Clarke, Experimental tests for the quantum
  behavior of a macroscopic degree of freedom: The phase difference across a
  josephson junction, {\em Phys. Rev. B} {\bf 35}, 4682 (Apr 1987).

\bibitem{koch2007charge}
J.~Koch, M.~Y. Terri, J.~Gambetta, A.~A. Houck, D.~I. Schuster, J.~Majer,
  A.~Blais, M.~H. Devoret, S.~M. Girvin and R.~J. Schoelkopf,
  Charge-insensitive qubit design derived from the cooper pair box, {\em
  Physical Review A} {\bf 76}, p. 042319  (2007).

\bibitem{martinis2002rabi}
J.~M. Martinis, S.~Nam, J.~Aumentado and C.~Urbina, Rabi oscillations in a
  large josephson-junction qubit, {\em Physical Review Letters} {\bf 89}, p.
  117901  (2002).

\bibitem{PhysRevA.57.120}
D.~Loss and D.~P. DiVincenzo, Quantum computation with quantum dots, {\em Phys.
  Rev. A} {\bf 57}, 120 (Jan 1998).

\bibitem{petta2005coherent}
J.~R. Petta, A.~C. Johnson, J.~M. Taylor, E.~A. Laird, A.~Yacoby, M.~D. Lukin,
  C.~M. Marcus, M.~P. Hanson and A.~C. Gossard, Coherent manipulation of
  coupled electron spins in semiconductor quantum dots, {\em Science} {\bf
  309}, 2180  (2005).

\bibitem{knill2001scheme}
E.~Knill, R.~Laflamme and G.~J. Milburn, A scheme for efficient quantum
  computation with linear optics, {\em Nature} {\bf 409}, 46  (2001).

\bibitem{jaksch1998cold}
D.~Jaksch, C.~Bruder, J.~I. Cirac, C.~W. Gardiner and P.~Zoller, Cold bosonic
  atoms in optical lattices, {\em Physical Review Letters} {\bf 81}, p. 3108
  (1998).

\bibitem{greiner2002quantum}
M.~Greiner, O.~Mandel, T.~Esslinger, T.~W. H{\"a}nsch and I.~Bloch, Quantum
  phase transition from a superfluid to a mott insulator in a gas of ultracold
  atoms, {\em Nature} {\bf 415}, 39  (2002).

\bibitem{lukin2001dipole}
M.~D. Lukin, M.~Fleischhauer, R.~Cote, L.~Duan, D.~Jaksch, J.~I. Cirac and
  P.~Zoller, Dipole blockade and quantum information processing in mesoscopic
  atomic ensembles, {\em Physical Review Letters} {\bf 87}, p. 037901  (2001).

\bibitem{gaetan2009observation}
A.~Ga{\"e}tan, Y.~Miroshnychenko, T.~Wilk, A.~Chotia, M.~Viteau, D.~Comparat,
  P.~Pillet, A.~Browaeys and P.~Grangier, Observation of collective excitation
  of two individual atoms in the rydberg blockade regime, {\em Nature Physics}
  {\bf 5}, 115  (2009).

\bibitem{saffman2010quantum}
M.~Saffman, T.~G. Walker and K.~M{\o}lmer, Quantum information with rydberg
  atoms, {\em Reviews of Modern Physics} {\bf 82}, p. 2313  (2010).

\bibitem{endres2016atom}
M.~Endres, H.~Bernien, A.~Keesling, H.~Levine, E.~R. Anschuetz, A.~Krajenbrink,
  C.~Senko, V.~Vuletic, M.~Greiner and M.~D. Lukin, Atom-by-atom assembly of
  defect-free one-dimensional cold atom arrays, {\em Science} {\bf 354}, 1024
  (2016).

\bibitem{bruzewicz2019trapped}
C.~D. Bruzewicz, J.~Chiaverini, R.~McConnell and J.~M. Sage, Trapped-ion
  quantum computing: Progress and challenges, {\em Applied Physics Reviews}
  {\bf 6}, p. 021314  (2019).

\bibitem{kjaergaard2020superconducting}
M.~Kjaergaard, M.~E. Schwartz, J.~Braum{\"u}ller, P.~Krantz, J.~I.-J. Wang,
  S.~Gustavsson and W.~D. Oliver, Superconducting qubits: Current state of
  play, {\em Annual Review of Condensed Matter Physics} {\bf 11}, 369  (2020).

\bibitem{arute2019quantum}
F.~Arute, K.~Arya, R.~Babbush, D.~Bacon, J.~C. Bardin, R.~Barends, R.~Biswas,
  S.~Boixo, F.~G. Brandao, D.~A. Buell {\em et~al.}, Quantum supremacy using a
  programmable superconducting processor, {\em Nature} {\bf 574}, 505  (2019).

\bibitem{wu2021strong}
Y.~Wu, W.-S. Bao, S.~Cao, F.~Chen, M.-C. Chen, X.~Chen, T.-H. Chung, H.~Deng,
  Y.~Du, D.~Fan {\em et~al.}, Strong quantum computational advantage using a
  superconducting quantum processor, {\em Physical review letters} {\bf 127},
  p. 180501  (2021).

\bibitem{zhu2022quantum}
Q.~Zhu, S.~Cao, F.~Chen, M.-C. Chen, X.~Chen, T.-H. Chung, H.~Deng, Y.~Du,
  D.~Fan, M.~Gong {\em et~al.}, Quantum computational advantage via 60-qubit
  24-cycle random circuit sampling, {\em Science bulletin} {\bf 67}, 240
  (2022).

\bibitem{pan2021solving}
F.~Pan, K.~Chen and P.~Zhang, Solving the sampling problem of the sycamore
  quantum supremacy circuits, {\em arXiv preprint arXiv:2111.03011}   (2021).

\bibitem{unruh1995maintaining}
W.~G. Unruh, Maintaining coherence in quantum computers, {\em Physical Review
  A} {\bf 51}, p. 992  (1995).

\bibitem{landauer1995quantum}
R.~Landauer, Is quantum mechanics useful?, {\em Philosophical Transactions of
  the Royal Society of London. Series A: Physical and Engineering Sciences}
  {\bf 353}, 367  (1995).

\bibitem{haroche1996quantum}
S.~Haroche and J.-M. Raimond, Quantum computing: dream or nightmare?, {\em
  Physics Today} {\bf 49}, 51  (1996).

\bibitem{shor1995scheme}
P.~W. Shor, Scheme for reducing decoherence in quantum computer memory, {\em
  Physical Review A} {\bf 52}, p. R2493  (1995).

\bibitem{steane1996error}
A.~M. Steane, Error correcting codes in quantum theory, {\em Physical Review
  Letters} {\bf 77}, p. 793  (1996).

\bibitem{knill1997theory}
E.~Knill and R.~Laflamme, Theory of quantum error-correcting codes, {\em
  Physical Review A} {\bf 55}, p. 900  (1997).

\bibitem{gottesman1997stabilizer}
D.~Gottesman, {\em Stabilizer codes and quantum error correction} (California
  Institute of Technology, 1997).

\bibitem{shor1996fault}
P.~W. Shor, Fault-tolerant quantum computation, in {\em Proceedings of 37th
  Conference on Foundations of Computer Science\/},  1996 pp. 56--65.

\bibitem{aharonov2008fault}
D.~Aharonov and M.~Ben-Or, Fault-tolerant quantum computation with constant
  error rate, {\em SIAM Journal on Computing}   (2008).

\bibitem{knill1998resilient}
E.~Knill, R.~Laflamme and W.~H. Zurek, Resilient quantum computation, {\em
  Science} {\bf 279}, 342  (1998).

\bibitem{kitaev1997quantum}
A.~Y. Kitaev, Quantum computations: algorithms and error correction, {\em
  Uspekhi Matematicheskikh Nauk} {\bf 52}, 53  (1997).

\bibitem{preskill1998reliable}
J.~Preskill, Reliable quantum computers, {\em Proceedings of the Royal Society
  of London. Series A: Mathematical, Physical and Engineering Sciences} {\bf
  454}, 385  (1998).

\bibitem{preskill1998fault}
J.~Preskill, Fault-tolerant quantum computation, in {\em Introduction to
  Quantum Computation and Information\/},  (World Scientific, 1998) pp.
  213--269.

\bibitem{aliferis2005quantum}
P.~Aliferis, D.~Gottesman and J.~Preskill, Quantum accuracy threshold for
  concatenated distance-3 codes, {\em Quantum Inf. Comput.} {\bf 6}, 97
  (2005).

\bibitem{reichardt2006fault}
B.~W. Reichardt, Fault-tolerance threshold for a distance-three quantum code,
  in {\em International Colloquium on Automata, Languages, and Programming\/},
  2006 pp. 50--61.

\bibitem{kitaev2003fault}
A.~Y. Kitaev, Fault-tolerant quantum computation by anyons, {\em Annals of
  Physics} {\bf 303}, 2  (2003).

\bibitem{dennis2002topological}
E.~Dennis, A.~Kitaev, A.~Landahl and J.~Preskill, Topological quantum memory,
  {\em Journal of Mathematical Physics} {\bf 43}, 4452  (2002).

\bibitem{raussendorf2007fault}
R.~Raussendorf and J.~Harrington, Fault-tolerant quantum computation with high
  threshold in two dimensions, {\em Physical Review Letters} {\bf 98}, p.
  190504  (2007).

\bibitem{fowler2012surface}
A.~G. Fowler, M.~Mariantoni, J.~M. Martinis and A.~N. Cleland, Surface codes:
  Towards practical large-scale quantum computation, {\em Physical Review A}
  {\bf 86}, p. 032324  (2012).

\bibitem{wen1990topological}
X.-G. Wen, Topological orders in rigid states, {\em International Journal of
  Modern Physics B} {\bf 4}, 239  (1990).

\bibitem{tsui1982two}
D.~C. Tsui, H.~L. Stormer and A.~C. Gossard, Two-dimensional magnetotransport
  in the extreme quantum limit, {\em Physical Review Letters} {\bf 48}, p. 1559
   (1982).

\bibitem{laughlin1983anomalous}
R.~B. Laughlin, Anomalous quantum hall effect: an incompressible quantum fluid
  with fractionally charged excitations, {\em Physical Review Letters} {\bf
  50}, p. 1395  (1983).

\bibitem{chen2010local}
X.~Chen, Z.-C. Gu and X.-G. Wen, Local unitary transformation, long-range
  quantum entanglement, wave function renormalization, and topological order,
  {\em Physical Review B} {\bf 82}, p. 155138  (2010).

\bibitem{haldane1983nonlinear}
F.~D.~M. Haldane, Nonlinear field theory of large-spin heisenberg
  antiferromagnets: semiclassically quantized solitons of the one-dimensional
  easy-axis n{\'e}el state, {\em Physical Review Letters} {\bf 50}, p. 1153
  (1983).

\bibitem{kane2005quantum}
C.~L. Kane and E.~J. Mele, Quantum spin hall effect in graphene, {\em Physical
  Review Letters} {\bf 95}, p. 226801  (2005).

\bibitem{chen2012symmetry}
X.~Chen, Z.-C. Gu, Z.-X. Liu and X.-G. Wen, Symmetry-protected topological
  orders in interacting bosonic systems, {\em Science} {\bf 338}, 1604  (2012).

\bibitem{bombelli1986quantum}
L.~Bombelli, R.~K. Koul, J.~Lee and R.~D. Sorkin, Quantum source of entropy for
  black holes, {\em Physical Review D} {\bf 34}, p. 373  (1986).

\bibitem{srednicki1993entropy}
M.~Srednicki, Entropy and area, {\em Physical Review Letters} {\bf 71}, p. 666
  (1993).

\bibitem{hastings2007area}
M.~B. Hastings, An area law for one-dimensional quantum systems, {\em Journal
  of statistical mechanics: theory and experiment} {\bf 2007}, p. P08024
  (2007).

\bibitem{white1992density}
S.~R. White, Density matrix formulation for quantum renormalization groups,
  {\em Physical Review Letters} {\bf 69}, p. 2863  (1992).

\bibitem{fannes1992finitely}
M.~Fannes, B.~Nachtergaele and R.~F. Werner, Finitely correlated states on
  quantum spin chains, {\em Communications in Mathematical Physics} {\bf 144},
  443  (1992).

\bibitem{vidal2004efficient}
G.~Vidal, Efficient simulation of one-dimensional quantum many-body systems,
  {\em Physical Review Letters} {\bf 93}, p. 040502  (2004).

\bibitem{verstraete2004renormalization}
F.~Verstraete and J.~I. Cirac, Renormalization algorithms for quantum-many body
  systems in two and higher dimensions, {\em arXiv preprint cond-mat/0407066}
  (2004).

\bibitem{kitaev2006topological}
A.~Kitaev and J.~Preskill, Topological entanglement entropy, {\em Physical
  Review Letters} {\bf 96}, p. 110404  (2006).

\bibitem{levin2006detecting}
M.~Levin and X.-G. Wen, Detecting topological order in a ground state wave
  function, {\em Physical Review Letters} {\bf 96}, p. 110405  (2006).

\bibitem{li2008entanglement}
H.~Li and F.~D.~M. Haldane, Entanglement spectrum as a generalization of
  entanglement entropy: Identification of topological order in non-abelian
  fractional quantum hall effect states, {\em Physical Review Letters} {\bf
  101}, p. 010504  (2008).

\bibitem{kitaev2002classical}
A.~Y. Kitaev, A.~Shen, M.~N. Vyalyi and M.~N. Vyalyi, {\em {Classical and
  Quantum Computation}}, no.~47 (American Mathematical Soc., 2002).

\bibitem{gottesman2009quantum}
D.~Gottesman and S.~Irani, The quantum and classical complexity of
  translationally invariant tiling and hamiltonian problems, in {\em 2009 50th
  Annual IEEE Symposium on Foundations of Computer Science\/},  2009 pp.
  95--104.

\bibitem{calabrese2005evolution}
P.~Calabrese and J.~Cardy, Evolution of entanglement entropy in one-dimensional
  systems, {\em Journal of Statistical Mechanics: Theory and Experiment} {\bf
  2005}, p. P04010  (2005).

\bibitem{lloyd1996universal}
S.~Lloyd, Universal quantum simulators, {\em Science} {\bf 273}, 1073  (1996).

\bibitem{Hawking1975}
S.~W. Hawking, Particle creation by black holes, {\em Comm. Math. Phys.} {\bf
  43}, 199  (1975).

\bibitem{bekenstein1974generalized}
J.~D. Bekenstein, Generalized second law of thermodynamics in black-hole
  physics, {\em Physical Review D} {\bf 9}, p. 3292  (1974).

\bibitem{israel1967event}
W.~Israel, Event horizons in static vacuum space-times, {\em Physical Review}
  {\bf 164}, p. 1776  (1967).

\bibitem{carter1971axisymmetric}
B.~Carter, Axisymmetric black hole has only two degrees of freedom, {\em
  Physical Review Letters} {\bf 26}, p. 331  (1971).

\bibitem{maldacena1999large}
J.~Maldacena, The large-{$N$} limit of superconformal field theories and
  supergravity, {\em International Journal of Theoretical Physics} {\bf 38},
  1113  (1999).

\bibitem{ryu2006holographic}
S.~Ryu and T.~Takayanagi, Holographic derivation of entanglement entropy from
  the anti--de sitter space/conformal field theory correspondence, {\em
  Physical Review Letters} {\bf 96}, p. 181602  (2006).

\bibitem{almheiri2015bulk}
A.~Almheiri, X.~Dong and D.~Harlow, Bulk locality and quantum error correction
  in ads/cft, {\em Journal of High Energy Physics} {\bf 2015}, 1  (2015).

\bibitem{pastawski2015holographic}
F.~Pastawski, B.~Yoshida, D.~Harlow and J.~Preskill, Holographic quantum
  error-correcting codes: Toy models for the bulk/boundary correspondence, {\em
  Journal of High Energy Physics} {\bf 2015}, 1  (2015).

\bibitem{hayden2007black}
P.~Hayden and J.~Preskill, Black holes as mirrors: quantum information in
  random subsystems, {\em Journal of High Energy Physics} {\bf 2007}, p. 120
  (2007).

\bibitem{sekino2008fast}
Y.~Sekino and L.~Susskind, Fast scramblers, {\em Journal of High Energy
  Physics} {\bf 2008}, p. 065  (2008).

\bibitem{shenker2014black}
S.~H. Shenker and D.~Stanford, Black holes and the butterfly effect, {\em
  Journal of High Energy Physics} {\bf 2014}, 1  (2014).

\bibitem{maldacena2016bound}
J.~Maldacena, S.~H. Shenker and D.~Stanford, A bound on chaos, {\em Journal of
  High Energy Physics} {\bf 2016}, 1  (2016).

\bibitem{cotler2017black}
J.~S. Cotler, G.~Gur-Ari, M.~Hanada, J.~Polchinski, P.~Saad, S.~H. Shenker,
  D.~Stanford, A.~Streicher and M.~Tezuka, Black holes and random matrices,
  {\em Journal of High Energy Physics} {\bf 2017}, 1  (2017).

\bibitem{almheiri2020replica}
A.~Almheiri, T.~Hartman, J.~Maldacena, E.~Shaghoulian and A.~Tajdini, Replica
  wormholes and the entropy of hawking radiation, {\em Journal of High Energy
  Physics} {\bf 2020}, 1  (2020).

\bibitem{penington2022replica}
G.~Penington, S.~H. Shenker, D.~Stanford and Z.~Yang, Replica wormholes and the
  black hole interior, {\em Journal of High Energy Physics} {\bf 2022}, 1
  (2022).

\bibitem{page1993information}
D.~N. Page, Information in black hole radiation, {\em Physical Review Letters}
  {\bf 71}, p. 3743  (1993).

\bibitem{harlow2013quantum}
D.~Harlow and P.~Hayden, Quantum computation vs. firewalls, {\em Journal of
  High Energy Physics} {\bf 2013}, 1  (2013).

\bibitem{preskill2018quantum}
J.~Preskill, Quantum computing in the {NISQ} era and beyond, {\em Quantum} {\bf
  2}, p.~79  (2018).

\bibitem{ai2021exponential}
{Google Quantum AI}, Exponential suppression of bit or phase errors with cyclic
  error correction, {\em Nature} {\bf 595}, p. 383  (2021).

\bibitem{ryan2021realization}
C.~Ryan-Anderson, J.~Bohnet, K.~Lee, D.~Gresh, A.~Hankin, J.~Gaebler,
  D.~Francois, A.~Chernoguzov, D.~Lucchetti, N.~Brown {\em et~al.}, Realization
  of real-time fault-tolerant quantum error correction, {\em Physical Review X}
  {\bf 11}, p. 041058  (2021).

\bibitem{egan2020fault}
L.~Egan, D.~M. Debroy, C.~Noel, A.~Risinger, D.~Zhu, D.~Biswas, M.~Newman,
  M.~Li, K.~R. Brown, M.~Cetina {\em et~al.}, Fault-tolerant operation of a
  quantum error-correction code, {\em arXiv preprint arXiv:2009.11482}
  (2020).

\bibitem{krinner2022realizing}
S.~Krinner, N.~Lacroix, A.~Remm, A.~Di~Paolo, E.~Genois, C.~Leroux,
  C.~Hellings, S.~Lazar, F.~Swiadek, J.~Herrmann {\em et~al.}, Realizing
  repeated quantum error correction in a distance-three surface code, {\em
  Nature} {\bf 605}, 669  (2022).

\bibitem{gidney2021factor}
C.~Gidney and M.~Eker{\aa}, How to factor 2048 bit {RSA} integers in 8 hours
  using 20 million noisy qubits, {\em Quantum} {\bf 5}, p. 433  (2021).

\bibitem{panteleev2022asymptotically}
P.~Panteleev and G.~Kalachev, Asymptotically good quantum and locally testable
  classical ldpc codes, in {\em Proceedings of the 54th Annual ACM SIGACT
  Symposium on Theory of Computing\/},  2022 pp. 375--388.

\bibitem{leverrier2022quantum}
A.~Leverrier and G.~Z{\'e}mor, Quantum tanner codes, {\em arXiv preprint
  arXiv:2202.13641}   (2022).

\bibitem{kitaev2001unpaired}
A.~Y. Kitaev, Unpaired majorana fermions in quantum wires, {\em
  Physics-{Uspekhi}} {\bf 44}, p. 131  (2001).

\bibitem{aghaee2022inas}
M.~Aghaee, A.~Akkala, Z.~Alam, R.~Ali, A.~A. Ramirez, M.~Andrzejczuk, A.~E.
  Antipov, M.~Astafev, B.~Bauer, J.~Becker {\em et~al.}, {InAs-Al} hybrid
  devices passing the topological gap protocol, {\em arXiv preprint
  arXiv:2207.02472}   (2022).

\bibitem{gottesman2001encoding}
D.~Gottesman, A.~Kitaev and J.~Preskill, Encoding a qubit in an oscillator,
  {\em Physical Review A} {\bf 64}, p. 012310  (2001).

\bibitem{campagne2020quantum}
P.~Campagne-Ibarcq, A.~Eickbusch, S.~Touzard, E.~Zalys-Geller, N.~E. Frattini,
  V.~V. Sivak, P.~Reinhold, S.~Puri, S.~Shankar, R.~J. Schoelkopf {\em et~al.},
  Quantum error correction of a qubit encoded in grid states of an oscillator,
  {\em Nature} {\bf 584}, 368  (2020).

\bibitem{fluhmann2019encoding}
C.~Fl{\"u}hmann, T.~L. Nguyen, M.~Marinelli, V.~Negnevitsky, K.~Mehta and
  J.~Home, Encoding a qubit in a trapped-ion mechanical oscillator, {\em
  Nature} {\bf 566}, 513  (2019).

\bibitem{puri2019stabilized}
S.~Puri, A.~Grimm, P.~Campagne-Ibarcq, A.~Eickbusch, K.~Noh, G.~Roberts,
  L.~Jiang, M.~Mirrahimi, M.~H. Devoret and S.~M. Girvin, Stabilized cat in a
  driven nonlinear cavity: a fault-tolerant error syndrome detector, {\em
  Physical Review X} {\bf 9}, p. 041009  (2019).

\bibitem{lescanne2020exponential}
R.~Lescanne, M.~Villiers, T.~Peronnin, A.~Sarlette, M.~Delbecq, B.~Huard,
  T.~Kontos, M.~Mirrahimi and Z.~Leghtas, Exponential suppression of bit-flips
  in a qubit encoded in an oscillator, {\em Nature Physics} {\bf 16}, 509
  (2020).

\bibitem{grimm2020stabilization}
A.~Grimm, N.~E. Frattini, S.~Puri, S.~O. Mundhada, S.~Touzard, M.~Mirrahimi,
  S.~M. Girvin, S.~Shankar and M.~H. Devoret, Stabilization and operation of a
  kerr-cat qubit, {\em Nature} {\bf 584}, 205  (2020).

\bibitem{nguyen2019high}
L.~B. Nguyen, Y.-H. Lin, A.~Somoroff, R.~Mencia, N.~Grabon and V.~E.
  Manucharyan, High-coherence fluxonium qubit, {\em Physical Review X} {\bf 9},
  p. 041041  (2019).

\bibitem{bao2022fluxonium}
F.~Bao, H.~Deng, D.~Ding, R.~Gao, X.~Gao, C.~Huang, X.~Jiang, H.-S. Ku, Z.~Li,
  X.~Ma {\em et~al.}, Fluxonium: an alternative qubit platform for
  high-fidelity operations, {\em Physical Review Letters} {\bf 129}, p. 010502
  (2022).

\bibitem{brooks2013protected}
P.~Brooks, A.~Kitaev and J.~Preskill, Protected gates for superconducting
  qubits, {\em Physical Review A} {\bf 87}, p. 052306  (2013).

\bibitem{gyenis2021experimental}
A.~Gyenis, P.~S. Mundada, A.~Di~Paolo, T.~M. Hazard, X.~You, D.~I. Schuster,
  J.~Koch, A.~Blais and A.~A. Houck, Experimental realization of a protected
  superconducting circuit derived from the 0--$\pi$ qubit, {\em PRX Quantum}
  {\bf 2}, p. 010339  (2021).

\bibitem{semeghini2021probing}
G.~Semeghini, H.~Levine, A.~Keesling, S.~Ebadi, T.~T. Wang, D.~Bluvstein,
  R.~Verresen, H.~Pichler, M.~Kalinowski, R.~Samajdar {\em et~al.}, Probing
  topological spin liquids on a programmable quantum simulator, {\em Science}
  {\bf 374}, 1242  (2021).

\bibitem{anderson1973resonating}
P.~W. Anderson, Resonating valence bonds: A new kind of insulator?, {\em
  Materials Research Bulletin} {\bf 8}, 153  (1973).

\bibitem{mi2022time}
X.~Mi, M.~Ippoliti, C.~Quintana, A.~Greene, Z.~Chen, J.~Gross, F.~Arute,
  K.~Arya, J.~Atalaya, R.~Babbush {\em et~al.}, Time-crystalline eigenstate
  order on a quantum processor, {\em Nature} {\bf 601}, 531  (2022).

\bibitem{wilczek2012quantum}
F.~Wilczek, Quantum time crystals, {\em Physical Review Letters} {\bf 109}, p.
  160401  (2012).

\bibitem{choi2017observation}
S.~Choi, J.~Choi, R.~Landig, G.~Kucsko, H.~Zhou, J.~Isoya, F.~Jelezko,
  S.~Onoda, H.~Sumiya, V.~Khemani {\em et~al.}, Observation of discrete
  time-crystalline order in a disordered dipolar many-body system, {\em Nature}
  {\bf 543}, 221  (2017).

\bibitem{zhang2017observation}
J.~Zhang, P.~W. Hess, A.~Kyprianidis, P.~Becker, A.~Lee, J.~Smith, G.~Pagano,
  I.-D. Potirniche, A.~C. Potter, A.~Vishwanath {\em et~al.}, Observation of a
  discrete time crystal, {\em Nature} {\bf 543}, 217  (2017).

\bibitem{kyprianidis2021observation}
A.~Kyprianidis, F.~Machado, W.~Morong, P.~Becker, K.~S. Collins, D.~V. Else,
  L.~Feng, P.~W. Hess, C.~Nayak, G.~Pagano {\em et~al.}, Observation of a
  prethermal discrete time crystal, {\em Science} {\bf 372}, 1192  (2021).

\bibitem{lee2022there}
S.~Lee, J.~Lee, H.~Zhai, Y.~Tong, A.~M. Dalzell, A.~Kumar, P.~Helms, J.~Gray,
  Z.-H. Cui, W.~Liu {\em et~al.}, Is there evidence for exponential quantum
  advantage in quantum chemistry?, {\em arXiv preprint arXiv:2208.02199}
  (2022).

\bibitem{jordan2012quantum}
S.~P. Jordan, K.~S. Lee and J.~Preskill, Quantum algorithms for quantum field
  theories, {\em Science} {\bf 336}, 1130  (2012).

\bibitem{gao2017traversable}
P.~Gao, D.~L. Jafferis and A.~C. Wall, Traversable wormholes via a double trace
  deformation, {\em Journal of High Energy Physics} {\bf 2017}, 1  (2017).

\bibitem{maldacena2017diving}
J.~Maldacena, D.~Stanford and Z.~Yang, Diving into traversable wormholes, {\em
  Fortschritte der Physik} {\bf 65}, p. 1700034  (2017).

\bibitem{brown2019quantum}
A.~R. Brown, H.~Gharibyan, S.~Leichenauer, H.~W. Lin, S.~Nezami, G.~Salton,
  L.~Susskind, B.~Swingle and M.~Walter, Quantum gravity in the lab:
  teleportation by size and traversable wormholes, {\em arXiv preprint
  arXiv:1911.06314}   (2019).

\bibitem{nezami2021quantum}
S.~Nezami, H.~W. Lin, A.~R. Brown, H.~Gharibyan, S.~Leichenauer, G.~Salton,
  L.~Susskind, B.~Swingle and M.~Walter, Quantum gravity in the lab:
  teleportation by size and traversable wormholes, part ii, {\em arXiv preprint
  arXiv:2102.01064}   (2021).

\bibitem{bernstein2017post}
D.~J. Bernstein and T.~Lange, Post-quantum cryptography, {\em Nature} {\bf
  549}, 188  (2017).

\bibitem{vazirani2019fully}
U.~Vazirani and T.~Vidick, Fully device independent quantum key distribution,
  {\em Communications of the ACM} {\bf 62}, 133  (2019).

\bibitem{wehner2018quantum}
S.~Wehner, D.~Elkouss and R.~Hanson, Quantum internet: A vision for the road
  ahead, {\em Science} {\bf 362}, p. eaam9288  (2018).

\bibitem{degen2017quantum}
C.~L. Degen, F.~Reinhard and P.~Cappellaro, Quantum sensing, {\em Reviews of
  Modern Physics} {\bf 89}, p. 035002  (2017).

\bibitem{mahadev2018classical}
U.~Mahadev, Classical verification of quantum computations, in {\em 2018 IEEE
  59th Annual Symposium on Foundations of Computer Science (FOCS)\/},  2018 pp.
  259--267.

\end{thebibliography}
\end{document}